# D-branes and Topological Charge in QCD

**H. B. Thacker***
*University of Virginia*
*E-mail:* `hbt8r@virginia.edu`

The recently observed long-range coherent structure of topological charge fluctuations in QCD is compared with theoretical expectations based on the AdS/CFT brane construction of nonsupersymmetric gauge theory by Witten. Similar observations of coherent topological charge structure in 2D $CP^{N-1}$ sigma models are interpreted in terms of Wilson lines representing world lines of screened electric charges. The analogy between 2D U(1) and 4D Yang-Mills theory leads to the interpretation of the observed coherent sheets of topological charge in QCD as screened "Wilson bags" first suggested by Luscher. The duality between the Wilson bag surface and a wrapped 6-brane in IIA string theory is discussed. The complete screening of the force between bag surfaces for integer values of the bag charge $\theta/2\pi$ corresponds to the observation by Polchinski that the net force between d-branes from closed string exchange vanishes for quantized values of Ramond-Ramond charge.

*Speaker.

## 1. D-branes and Theta Dependence

AdS/CFT string/gauge duality provides new insight into some longstanding issues in QCD. It is particularly well suited to the discussion of long-range nonperturbative structure. Thus, for example, in this framework the QCD flux tube associated with quark confinement in $SU(N)$ gauge theory is interpreted as the holographic image of a fundamental type IIA string. In Witten's brane construction [1], 4D SU(N) gauge theory is obtained from type IIA string theory on $R_4 \times S_1 \times R_5$ with N coincident 4-branes filling the $R_4 \times S_1$. The boundary conditions on the $S_1$ break the supersymmetry, and the gravitational field of the 4-branes forms a 5-dimensional black hole in the dimensions transverse to the 4-branes. The string theory in this metric is holographically equivalent to a gauge theory on the horizon whose long distance behavior (long compared to the radius of the compactified dimension) is believed to be in the same universality class as ordinary nonsupersymmetric QCD.

One of the most interesting implications of string/gauge duality for QCD is in the nature of theta dependence and topological charge structure [2]. The theta term in 4D QCD, of the form $\theta F \wedge F$, arises from a 5D Chern-Simons term $a \wedge F \wedge F$ defined on the world volume of the 4-branes. Here $a_\mu$ is the abelian gauge field that couples to Ramond-Ramond (RR) charge in IIA string theory. Thus QCD topological charge is holographically dual to RR charge, which is a solitonic ("magnetic") charge whose existence is implied by string theory duality arguments but which is not carried by ordinary string states. The seminal results of Polchinski [3] demonstrated that RR charge in fact resides on D-branes. In Witten's construction, the geometry induced by the 4-branes is $R_4 \times D \times S_4$, where D is a 2-dimensional disk with a Schwarzschild singularity at its center. The QCD $\theta$ parameter is equal (mod $2\pi$) to the Wilson line integral of the RR gauge field around the $S_1$, which forms the circumference of the disk. (The $S_4$ plays a passive role, except to serve as a place to wrap 4 of the 6 spatial directions of 6-branes, which play a central role in the discussion of topological charge.) To clarify the nature of this construction, we may consider a similar holographic view of 2-dimensional U(1) gauge theories like $CP^{N-1}$. At a fixed time, we represent the spatial axis by a long, thin, solid cylinder ($R_1 \to R_1 \times D$), as depicted in Fig. 1. This figure brings out the interesting fact that Witten's holographic view of theta-dependence, when applied to a 2D U(1) gauge theory like $CP^{N-1}$, is in fact identical to Laughlin's famous gedankenexperiment which demonstrated the topological origin of the integer quantum Hall effect.

The basic picture of topological charge structure that emerges from the AdS/CFT correspondence was actually proposed two decades earlier by both Luscher [4] and Witten [5]. In this picture, there are multiple "k-vacuum" states which are labeled by an integer $k$, where the local effective value of theta is shifted from its background value by $2\pi k$ (corresponding in the string theory to the number of units of RR flux threaded through the singularity). Different k-vacua will be separated by domain walls. In the string theory, these domain walls correspond to D6-branes which are wrapped around the $S_4$, and which therefore appear in 3+1 D as 2+1-dimensional objects, i.e. domain walls or membranes. The fact that the local value of the line integral that defines $\theta$ jumps by $\pm 2\pi$ when crossing a domain wall reflects the quantization of RR charge on a D-brane. This k-vacuum scenario is in some sense a 4D generalization of the original discussion of theta-dependence in the 2D massive Schwinger model [6]. In that case, $\theta$ can be interpreted as a background electric field, and a domain wall is just a charged particle, which separates two spatial regions in which the electric

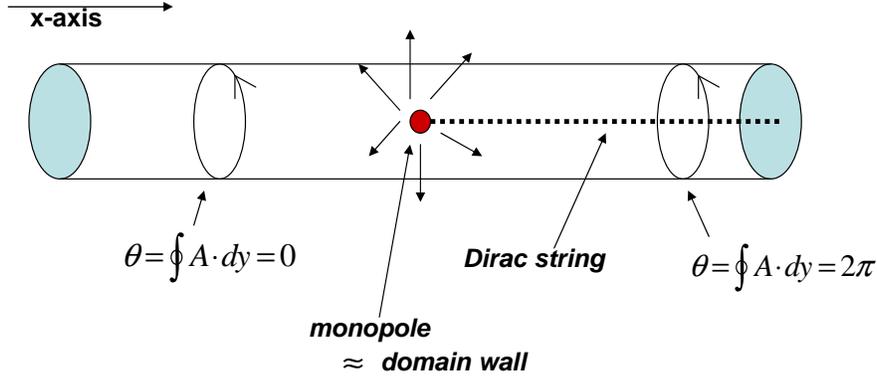

**Figure 1:** Holographic view of a domain wall in (1+1)-dimensional $CP^{N-1}$ theories from a (3+1)-dimensional perspective. The long axis of the cylinder becomes the spatial axis of the (1+1)-dimensional theory. Plot is at a fixed time.

flux density differs by one unit.

Luscher's discussion draws on the analogy between 4D QCD and 2D $CP^{N-1}$ models. He pointed out that in both $CP^{N-1}$ and QCD, finite topological susceptibility implies the presence of a massless pole in the correlator of two Chern-Simons currents. leading to a "hidden long-range order." This does not imply a massless particle in either theory because the CS current is non-gauge-invariant, but the residue of the massless pole is the gauge invariant topological susceptibility.

In 2D U(1), the CS current is just the dual of the gauge potential

$$j_\mu^{CS} = \varepsilon_{\mu\nu} A^\nu \qquad (1.1)$$

and the massless pole corresponds to a confining Coulomb potential between test U(1) charges and hence an area law for fractionally charged Wilson loops. In the $CP^{N-1}$ models, however, the gauge field is an auxiliary field without a kinetic term. The coulomb potential and finite topological susceptibility are generated by closed z-loops. (Here I use z to denote the $CP^{N-1}$ field.) The $A_\mu$ field satisfies an exact London equation which equates it to the current of charged z-particles,

$$A_\mu = \frac{1}{2} i \left( \mathbf{z}^\dagger \partial_\mu \mathbf{z} - (\partial_\mu \mathbf{z})^\dagger \mathbf{z} \right) \qquad (1.2)$$

This impies the complete screening of the Coulomb field of an integer-valued test charge. In two dimensions, placing a Wilson loop of charge $\theta/2\pi$ around a region of space-time is equivalent to adding a $\theta$ term to the action inside that region. The coefficient of the area law (string tension) is just the difference in vacuum energy density $E(\theta) - E(0)$. For a unit charged loop, $\theta = 2\pi$ and the area term vanishes by periodicity of $E(\theta)$. The $\theta = 2\pi$ vacuum inside the loop is essentially the $\theta = 0$ vacuum with a current flowing around its boundary. The basic quasipartcle excitation is thus a screened Wilson line representing a charged particle whose Coulomb field has been completely screened by the backflow of $z$-particles in the vacuum. The ground state of the model can be described as a condensate of such quasiparticles.

In 4D QCD the analog of the Wilson loop in 2D is not a Wilson loop but a "Wilson bag," i.e. the integral over a 3-surface of the Chern-Simons tensor,

$$A_{\mu\nu\rho} = -Tr\left(B_\mu B_\nu B_\rho + \frac{3}{2}B_{[\mu}\partial_\nu B_{\rho]}\right) \quad (1.3)$$

where $B_\mu$ is the Yang-Mills gauge potential. Just as the Wilson loop in 2D represents the world line of a charge particle, a Wilson bag can be interpreted as the world volume of a 2-space-dimensional charged membrane. Like the Wilson loop in 2D, the Wilson bag surface separates two vacua for which the values of $\theta$ differ by $2\pi$ (for a unit charged bag). It is clear that this Wilson bag surface is the gauge theory dual of the wrapped 6-brane in IIA string theory. Equivalently, it is a domain wall between k-vacua. Like the Wilson loop in $CP^{N-1}$, an integer-charged Wilson bag will be completely screened. For a fractionally charged bag, there would be a confining force between opposite walls of the bag, but for an integer charge, this force vanishes. Circumstantial evidence suggests that the string theory dual analog of this complete screening is the result shown by Polchinski [3] that the force between two D-branes due to closed string exchange vanishes (i.e. they are BPS states). This BPS property is associated with the quantized value of RR charge on the D-brane, which leads to an exact cancellation between NS-NS and R-R closed string exchange.

## 2. Evidence for D-branes in Monte Carlo Gauge Configurations

The advent of exactly chiral Ginsparg-Wilson fermions has provided a new and powerful tool for studying topological structure. If we construct an exactly chiral lattice Dirac operator $D$ (here we use the overlap operator [7]), the topological charge density operator

$$q(x) = \frac{g^2}{48\pi^2}F\tilde{F} \quad (2.1)$$

may be represented on the lattice entirely in terms of $D$,

$$q(x) = -\frac{1}{2a}tr\gamma_5 D(x,x) \quad (2.2)$$

where the trace is over spin and color indices. This definition of the topological charge density arises from the non-invariance of the measure in the calculation of the chiral anomaly on the lattice. This close connection to lattice chiral symmetry leads to a much smoother and less singular distribution of q(x) than that found with a more naive ultralocal definition of $q(x)$ constructed directly from the gauge links. As is well-known, such direct constructions of $q(x)$ lead to distributions with a large amount of short-range noise, which obscures any possible long-range structure that may be present. Cooling or ad-hoc smoothing procedures, which may be adequate for measuring the global charge of a configuration, are likely to produce misleading results for local distributions. By contrast, for both QCD and $CP^{N-1}$ models, the overlap-based topological charge operator is found to reveal the long range structure in the TC distributions without any cooling or smoothing.

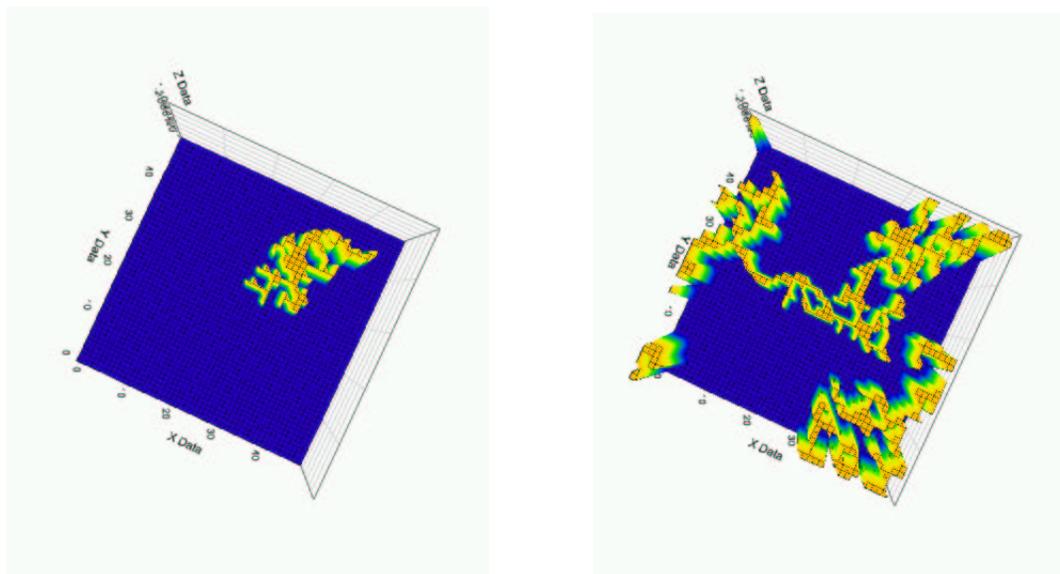

**Figure 2:** Largest structure for a typical random $q(x)$ distribution (left) compared with the largest structure for a typical ($\beta = 1.2$) $CP^3$ Monte Carlo configuration (right) on a $50 \times 50$ lattice.

The first study of topological charge distributions in 4D pure-glue QCD using the overlap operator was reported in [8]. This study produced a stunning result. It was found that, in each of the configurations studied, the TC distribution was dominated by two opposite-sign coherent connected structures which spread over the entire lattice (which was about $(1.5 \text{ fm})^4$). These structures occupy effectively 3-dimensional world volumes in the sense that they are everywhere thin in one of the four directions. The positive and negative charge structures are everywhere close together and folded or crumpled to occupy most of spacetime (typically about 80% of the lattice points). The two oppositely charged structures could thus be reasonably described as a single structure consisting of an extended dipole layer. The dimensionality of this layer is that of the world volume of a 2-brane or membrane. Further study of these structures [9] showed that they are inherently global, with the topological charge being distributed more or less uniformly throughout the coherent regions, as opposed to being concentrated in localized lumps. By studying these coherent structures for three different lattice spacings, it was also found that the thickness of the membranes scales to zero in the continuum limit. In fact, for finite lattice spacing, the thickness of the membranes was typically only one or two lattice spacings.

In view of the theoretical arguments outlined in the previous section, it is tempting to interpret the coherent dipole sheets of topological charge observed in the Monte Carlo configurations as Wilson bag surfaces, or equivalently, as the holographic image of wrapped 6-branes in Witten's construction. If this interpretation has any validity, we would expect, by Luscher's arguments, to find analogous 1-dimensional coherent structures in the 2D $CP^{N-1}$ model, corresponding to Wilson line excitations. Note that both the Wilson line and the Wilson bag excitations will produce a dipole layer of topological charge. In the 2D case, the topological charge is proportional to $\varepsilon_{\mu\nu}\partial^\mu A^\nu$, the derivative of $A_\mu$ transverse to its direction. If $A_\mu$ is a uniform constant along the Wilson line and zero everywhere else, this derivative produces two opposite sign 1D coherent regions which together form a dipole layer (coming from the derivative of a delta-function in the transverse coordinate). Similarly in 4D QCD, for an excitation of the Chern-Simons tensor which is uniform within the 3-dimensional world volume of the brane and zero elsewhere, the topological charge is given by it's derivative transverse to the brane. This will produce a structure with positive and negative coherent regions on opposite sides of the bag surface, forming a dipole membrane, much like what is seen in the Monte Carlo studies.

The essential properties of the observed coherent topological charge strucure can be directly related to expected qualitative features of the two-point correlator $G(x) = \langle q(x)q(0) \rangle$. This correlator consists of a zero-range positive contact term at $x = 0$, and a short-range negative tail for $x > 0$. The presence of coherent membranes of topological charge is clearly responsible for building up the positive contact term, which tends to explain why their thickness scales to zero. The alternating-sign layering of the vacuum is the mechanism for respecting the required negativity of the $x \neq 0$ correlator even though the charge distributions are dominated by large sign-coherent (but, crucially, lower-dimensional) regions.

The results of a Monte Carlo study of $CP^{N-1}$ models will be presented in detail elsewhere [10]. The overlap-based TC distributions were studied in detail for $CP^3$, revealing large-scale coherent connected regions which are essentially one-dimensional and arranged in alternating-sign layers. This is exactly what would be expected from coherent Wilson-line excitations and is precisely the analog of the 4D QCD results. To illustrate the basic observation of topological charge structure in $CP^3$, Figure 2 compares the largest coherent connected structure on a randomly generated $q(x)$ distribution with one obtained on a typical $CP^3$ configuration on a $50 \times 50$ lattice at $\beta = 1.2$ (correlation length $\approx 19$). The "stringy" appearance of the $CP^3$ structures is quite obvious and provides visible support for the assertion of extended coherent 1D structure analogous to that found in QCD. As the nature of this topological charge structure is explored further, the $CP^{N-1}$ models should be very useful. These models are computationally and theoretically more easily studied than QCD, and it is encouraging to see that the analogy with QCD extends in detail to the dynamical structure of topological charge.

## References

[1] E. Witten, Adv. Theor. Mat. Phys. 2:253 (1998).

[2] E. Witten, Phys. Rev. Lett. 81: 2862 (1998).

[3] J. Polchinski, Phys. Rev. Lett. 75: 4724 (1995).

[4] M. Luscher, Phys. Lett. B78: 465 (1978).

[5] E. Witten, Nucl. Phys. B145: 110 (1978).

[6] S. Coleman, Ann. Phys. 101: 239 (1976).

[7] H. Neuberger, Phys. Lett. B417: 141 (1998).

[8] I. Horvath, et al, Phys. Rev. D68: 114505 (2003).

[9] I. Horvath, et al. Phys. Lett. B612: 21 (2005).

[10] S. Ahmad, J. Lenaghan, and H. Thacker (unpublished)